\documentstyle[12pt,epsf]{article}

\newlength{\pubnumber} 

\catcode`\@=11
\@addtoreset{equation}{section}

\def\section{\@startsection{section}{1}{\z@}{3.5ex plus 1ex minus .2ex}
 {2.3ex plus .2ex}{\large\bf}}
\def\subsection{\@startsection{subsection}{2}{\z@}{2.3ex plus .2ex}
 {2.3ex plus .2ex}{\bf}}

\def\beq{\begin{equation}}
\def\eeq{\end{equation}}
\def\beqn{\begin{eqnarray}}
\def\eeqn{\end{eqnarray}}

\def\half{{\textstyle{1\over 2}}}

\def\sixth{{\textstyle {1\over6}}}

\def\Tr{{\rm Tr}\, }

\def\vev#1{\langle #1\rangle}

\def\eps{\epsilon}

\def\p4{\Phi_4}
\def\pp4{\Phi^{'}_4}
\def\pb4{\bar{\Phi}_4}
\def\ppb4{\bar{\Phi}^{'}_4}
\def\p#1{{\Phi_{#1}}}
\def\P#1{{\Phi_{#1}}}
\def\pp#1{{\Phi^{'}_{#1}}}
\def\pb#1{{{\overline{\Phi}}_{#1}}}

\def\ppb#1{{{\overline{\Phi}}^{'}_{#1}}}

\def\FD2pv{FD2$^{'}$V }
\def\FD2p{FD2$^{'}$ }

\def\inbar{\,\vrule height1.5ex width.4pt depth0pt}

\def\IC{\relax\hbox{$\inbar\kern-.3em{\rm C}$}}
\def\IQ{\relax\hbox{$\inbar\kern-.3em{\rm Q}$}}
\def\IR{\relax{\rm I\kern-.18em R}}
 \font\cmss=cmss10 \font\cmsss=cmss10 at 7pt

\def\IZ{\relax\ifmmode\mathchoice
 {\hbox{\cmss Z\kern-.4em Z}}{\hbox{\cmss Z\kern-.4em Z}}
 {\lower.9pt\hbox{\cmsss Z\kern-.4em Z}}
 {\lower1.2pt\hbox{\cmsss Z\kern-.4em Z}}\else{\cmss Z\kern-.4em Z}\fi}

\def\Io{\relax\ifmmode\mathchoice
 {\hbox{\cmss 1\kern-.4em 1}}{\hbox{\cmss 1\kern-.4em 1}}
 {\lower.9pt\hbox{\cmsss 1\kern-.4em 1}}
 {\lower1.2pt\hbox{\cmsss 1\kern-.4em 1}}\else{\cmss 1\kern-.4em 1}\fi}

\def\u{\underline{\phantom{a}}}

\begin{document}

\begin{titlepage}
\samepage{
\setcounter{page}{1}
\rightline{BU-HEPP-03/02}
\rightline{CASPER-03/17}
\rightline{\tt hep-ph/0310155}
\rightline{September 2003}
\vfill
\begin{center}
 {\Large \bf  Heterotic String Optical Unification\\
                    }
\vfill
\vskip .4truecm
\vfill {\large
        J. Perkins,$^{1}$\footnote{John$\u$Perkins@baylor.edu} 
        B. Dundee,$^{1}$\footnote{Ben$\u$Dundee@baylor.edu} 
        R. Obousy,$^{1}$\footnote{Richard$\u$K$\u$Obousy@baylor.edu}
        E. Kasper,$^{1,3}$\footnote{ekasper@physics.tamu.edu}
        M. Robinson,$^{1,4}$\footnote{robinmb@auburn.edu}
        K. Stone,$^{1,5}$\footnote{zkrs18@imail.etsu.edu}  
     \& G. Cleaver,$^{1,2}$\footnote{Gerald$\u$Cleaver@baylor.edu}
}
\\
\vspace{.12in}
{\it $^{1}$ Center for Astrophysics, Space Physics \& Engineering Research,\\
            Dept.\ of Physics, PO Box 97316, Baylor University,\\
            Waco, TX 76798-7316\\}
\vspace{.06in}
{\it $^{2}$ Astro Particle Physics Group,
            Houston Advanced Research Center (HARC),\\
            The Mitchell Campus,
            Woodlands, TX 77381\\}
\vspace{.06in}
{\it $^{3}$ Department of Physics,
            Texas A \& M University,\\
            College Station, TX 77843-4242\\}
\vspace{.06in}
{\it $^{4}$ Department of Physics,
            Auburn University,\\
            Auburn, AL 36849\\}
\vspace{.06in}
{\it $^{5}$ Department of Physics,
            East Tennessee State University,\\
            Johnson City, TN, 37614\\}
\vspace{.025in}
\end{center}
\vfill
\newpage
\begin{abstract}
The lower limit to string
coupling unification in weakly coupled heterotic strings 
was shown by Kaplunovsky to be around
$\Lambda_H \sim 5 \times 10^{17}$ GeV.
In contrast, under the scenario of an intermediate scale desert,
the $SU(3)_C\times SU(2)_L\times U(1)_Y$ ([321])
Minimal Supersymmetric Standard Model (MSSM) unification scale is 
$\Lambda_U \sim 2.5 \times 10^{16}$ GeV.
Optical unification was proposed by Giedt as a robust
mechanism for relating these two scales.
In this mechanism, intermediate scale MSSM-charged exotic particles 
effect running couplings like a diverging lens, always producing a ``virtual'' 
image of the string unification point between the string scale and the
exotic particle mass scale.  A heterotic string model
that offers the potential to realize optical unification  
was recently constructed.
This paper reports the initial results of an
investigation into the possibility of 
optical unification realization in this model.

{\it This paper is a product of the NSF REU  
program at Baylor University and is based on a talk presented at 
the Second International Conference on String Phenomenology, Durham, England, 
29 July - 4 August, 2003.} 
\end{abstract}
\smallskip}
\end{titlepage}

\setcounter{footnote}{0}


\section{Review of Optical Unification}

In weakly coupled heterotic strings,
the unification scale for running couplings has a lower limit of around
$\Lambda_H \sim 5\times 10^{17}$ GeV \cite{kapl}.  In contrast, 
the running of the couplings upward from measured values at $M_{Z^o}$ predict 
Minimal Supersymmetric Standard Model (MSSM)
$SU(3)_C\times SU(2)_L\times U(1)_Y$ ([321])
gauge coupling unification at $\Lambda_U\sim 2.5 \times 10^{16}$ GeV 
when an intermediate scale desert is assumed \cite{mssm}. 
One resolution is a grand unified theory (GUT) 
between the MSSM and string scales. However,
with the exception of flipped $SU(5)$ \cite{fsu5}
(or partial GUTs such as the Pati-Salam $SU(4)_C\times SU(2)_L\times SU(2)_R$ 
\cite{lrs,cfs}), 
string GUTs cannot be generated by level-one Ka\v c-Moody algebras 
(since they lack the required 
adjoint higgs and/or higher dimensional scalar representations) 
and models based on higher level Ka\v c-Moody algebras vastly prefer even 
numbers of generations \cite{hlguta,hlgutb,hlgutc}.
Alternately, strong coupling effects of
$M$-theory can lower $\Lambda_H$ down to $\Lambda_U$ \cite{witten}.
On the other hand, intermediate scale $\Lambda_I< \Lambda_U$ 
MSSM-charged exotics could shift
the MSSM unification scale upward to the string scale \cite{mup}.

The near ubiquitous appearance of MSSM-charged exotics in
heterotic string models adds weight to the third
proposal.  However, most intermediate scale MSSM-charged 
exotic solutions might be viewed as accidental.  While existence of 
the string unification scale would clearly be stable under shifts 
in the masses of the exotics, the prediction of an apparent lower-scale 
MSSM unification
when MSSM-charged exotics are ignored would generally be unstable under 
these mass shifts.  On the other hand, a set of intermediate scale MSSM-charged
exotic particles satisfying ``optical unification'' constraints provide
a robust method for stablizing apparent MSSM unification  
under such shifts \cite{jg1}. 
With optical unification, a shift of the intermediate scale $\Lambda_I$
produces a shift in $\Lambda_U$, rather than the disappearance
of $\Lambda_U$.
This effect is parallel to a diverging lens always producing 
a virtual image between itself and a real object,
independent of the position of the lens or real object.
Hence, the choice of appellation for this mechanism.

Successful optical unification requires three things \cite{jg1}.
First, the effective level of the
hypercharge generator must be the standard
\beqn
k_Y = \textstyle{5\over3}\, .
\label{db123k}
\eeqn
(\ref{db123k}) is a
strong constraint on string-derived $[321]$ models,
for the vast majority have non-standard
hypercharge levels.
Only select classes of models, such as the 
free fermionic \cite{fff} NAHE-based \cite{nahe} class,
can yield $k_{Y} = \textstyle{5\over3}$.

Second, optical unification imposes the relationship
\beqn
\delta b_2 = \textstyle{7\over12} \delta b_3 + \textstyle{1\over4} \delta b_Y\, .
\label{db123}
\eeqn
between the exotic particle contributions $\delta b_3$, $\delta b_2$, 
and $\delta b_1$
to the [321] beta function coefficients.
Each $SU(3)_C$ exotic triplet or anti-triplet contributes
$\half$ to $\delta b_3$;
each $SU(2)_C$ exotic doublet contributes
$\half$ to $\delta b_2$.
With the hypercharge of a MSSM quark doublet normalized to $\sixth$,
the contribution to $\delta b_Y$ from an individual particle with
hypercharge $Q_Y$ is $Q_Y^2$.
$\delta b_3 > \delta b_2$ is required
to keep the virtual unification scale below the string scale.
Combining all of this with (\ref{db123}) imposes
\beqn
\delta b_3 >  \delta b_2 \ge \textstyle{7\over12} \delta b_3 \, ,
\label{db123b}
\eeqn
since $\delta b_Y \geq 0$.

To acquire intermediate scale mass,
the exotic triplets and anti-triplets must be equal in number.
Similarly, there must be an even number of exotic doublets.
Hence, $\delta b_3$ and $\delta b_2$ must be integer.
The simplest solution to
(\ref{db123}) and (\ref{db123b}) is a set of
three exotic triplet/anti-triplet pairs and two pairs of doublets.
One pair of doublets can carry $Q_Y=\pm \half$, while the remaining
exotics carry no hypercharge \cite{jg1}.
Alternately, if the doublets carry too little hypercharge,
some exotic $SU(3)_C \times SU(2)_L$ singlets could make up the
hypercharge deficit.
The next simplest
solution requires four triplet/anti-triplet pairs and three pairs of
doublets that yield $\delta b_Y = 2 \textstyle{2\over 3}$
either as a set, or with the assistance of additional non-Abelian singlets.
For more than four triplet/anti-triplet pairs, (\ref{db123})
and (\ref{db123b}) allow varying numbers of pairs of doublets.

\section{Flat Directions of String Models}

Quasi-realistic heterotic string models generically contain an anomalous $U(1)_A$
(for which ${\rm Tr}\, Q^{(A)}\ne 0$) \cite{anomu1}. 
The Green-Schwarz-Dine-Seiberg-Witten mechanism \cite{gsdsw} breaks
the anomalous U(1), and in the process generates 
a Fayet-Iliopoulos (FI) term,
\beqn
\eps\equiv\frac{g^2_s M_P^2}{192\pi^2}\Tr Q^{(A)},
\label{fit}
\eeqn
in the associated $D$-term.
The FI term breaks supersymmetry near the Planck scale 
and destabilize the string vacuum, unless it is cancelled 
by scalar vacuum expectation values (VEVs),
\beqn
\vev{D_{A}} &=& \sum_m Q^{(A)}_m |\vev{\varphi_{m}}|^2 +
\eps  = 0\,\, . \label{daf}
\eeqn
Thus, an effect of the anomalous $U(1)_A$ is to 
induce a non-perturbatively chosen flat direction of VEVs. 
Since the fields taking on the VEVs
typically carry additional non-anomalous charges, a
non-trivial set of constraints is imposed on the VEVs.
The VEVs must maintain $D$-flatness for each non-anomalous gauge symmetry. 
For non-anomalous Abelian symmetries it is required that, 
\beqn
\vev{D_i} &=& \sum_m Q^{(i)}_m |\vev{\varphi_{m}}|^2 = 0\,\, ,
\label{dana}
\eeqn
while for non-Abelian symmetries,
\beqn
\vev{D_a^{\alpha}}&=& \sum_m
\vev{\varphi_{m}^{\dagger} T^{\alpha}_a \varphi_m} = 0\,\, ,
\label{dtgen}
\eeqn
with $T^{\alpha}_a$ a matrix generator of the non-Abelian symmetry.
The scalar VEVs will in general break some, or all, of the additional gauge
symmetries spontaneously.

To insure a
supersymmetric vacuum $F$-flatness,
\beq
\vev{F_{m}} \equiv \vev{\frac{\partial W}{\partial \Phi_{m}}} = 0,
\label{ff}
\eeq
must also be maintained for 
each superfield $\Phi_{m}$ (containing a scalar field $\varphi_{m}$
and chiral spin-$\half$ superpartner $\psi_m$) appearing
in the superpotential $W$.

A typical string model contains 
a moduli space of perturbative solutions to the $D$- and $F$-flatness constraints,
which are supersymmetric and degenerate in energy \cite{moduli}. 
Much of the study of the phenomenology of superstring models 
involves the analysis and classification of these flat directions.
The methods for this analysis in string models have been systematized
in recent years \cite{gc98,mshsm,hetdisc,cfs}.

\section{String Model with Optical Unification Potential} 

The possibility of optical unification
within the context of free fermionic heterotic strings was investigated 
recently \cite{gc1, gcps}.  
By altering some of the GSO projection coefficients 
in the models originally presented in \cite{af3}, 
but keeping the same free fermionic boundary vectors,
a model with a set of $SU(3)_C$ exotic triplet/anti-triplet pairs, $SU(2)_L$ 
exotic doublets, and non-Abelian singlets  
satisfying optical unification requirements was constructed. 
Specifically, the model contains three pairs 
of exotic triplet/anti-triplets with $Q_Y= \pm \frac{1}{3}$, 
one pair of exotic triplet/anti-triplets with $Q_Y= \pm \frac{1}{6}$, 
three pairs of exotic doublets with $Q_Y = 0$, and 
seven pairs of singlets with $Q_Y=\pm \frac{1}{2}$. 
The model also contains four copies of higgs doublet pairs. 
(See Table 1 below. The complete set of states are given in Ref.\ 
\cite{gc1}.) 
Optical unification requires that all four triplet/anti-triplet 
pairs, all three
doublet pairs, and one pair of of exotic singlets with hypercharge 
take on an intermediate scale mass. 
The remaining six pairs of exotic singlets with hypercharge must take on 
masses above the MSSM unification scale, 
as must three out of four orthogonal combinations of the higgs.

The proposed optical unification model contains an anomalous $U(1)_A$.  
Thus some of its scalar fields will necessarily receive FI-scale VEVs.  
We have begun a systematic search for flat direction VEVs that will
induce FI-scale masses for the 
six extra pairs of exotic $Q_Y=\pm \frac{1}{2}$ singlets and 
the three extra higgs,
while keeping the four exotic triplet/anti-triplets pairs, 
three exotic doublet pairs, and one  
exotic $Q_Y=\pm \frac{1}{2}$ singlent pair FI-scale massless.
The first step in our sytematic search for $D$- and $F$-flat directions 
was to generate a basis set of $D$-flat directions.   
Routines have been developed for this 
using a singular value decomposition approach \cite{mshsm,svg}.  

In our initial investigation only singlet fields with $Q_Y=0$
were allowed VEVs. 
The model contains 27 such fields: the three uncharged moduli
$\Phi_{i=1,2,3}$; $\Phi_{12}$, $\Phi_{23}$, $\Phi_{31}$, 
and their respective complex conjugate fields 
${\bar\Phi}_{12}$, ${\bar\Phi}_{23}$, ${\bar\Phi}_{31}$;
and $S_{j=1\, {\rm to }\, 9}$ 
and their respective complex conjugate fields, 
${\bar S}_{j=1\, {\rm to }\, 9}$.  
Of these, only $S_{7}$, $S_{8}$, and $S_{9}$ 
(and $\bar{S}_{7}$, $\bar{S}_{8}$, and $\bar{S}_{9}$) 
carry anomalous charge.
All $D$-flat directions containing only VEVs of these 27 singlet 
fields can be generated from a basis set of nine elements, denoted
${\cal D}_{i=1\, {\rm to}\, 9}$, given in Table 3 below. 
Only ${\cal D}_1$, corresponds to a net negative anomalous charge,
while all of the other eight correspond to no anomalous charge. 
Since the FI term has positive sign, any $D$-flat direction must contain 
${\cal D}_1$ and can contain any number of the ${\cal D}_{i>1}$.

We applied stringent $F$-flatness constraints to all $D$-flat directions 
formed as 
linear combinations of the basis elements ${\cal D}_{i}$, 
with coefficients ranging from 1 to 50 for ${\cal D}_1$ 
and ${\cal D}_{i\ge 7}$, 
and ranging from -50 to 50 for ${\cal D}_{1< i < 7}$. 
Since the $\Phi_{12}$, $\Phi_{23}$, $\Phi_{31}$, and $S_i$ fields 
have vector-like partner (complex conjugate) fields,
a negative coefficient implies that the relevant 
vector partners take on the net VEVs.     
Stringent $F$-flat directions are those for which $F$-flatness does not 
require cancellation between two or more terms in a given $F$-term.
This implies that at least two fields in a given $F$-term do not take on VEVs. 
We found that only one class of $D$-flat directions was 
stringently $F$-flat to all order. The FI-scale 
field VEVs for this class are: 
\beqn
<{\bar S}_7 {\bar S}_8 S_9> <{\bar \Phi}_{12}\, {\rm and/or}\,
                             {\bar \Phi}_{31}\, {\rm and/or}\, 
                             {\Phi}_2>\, .
\label{sfset}
\eeqn
That is, the fields ${\bar S}_7$, ${\bar S}_8$, and $S_9$ must take on VEVs,
while VEVs are optional for ${\bar\Phi}_{12}$, ${\bar \Phi}_{31}$, and 
$\Phi_2$. The relative ratios of the squares of the norms of the VEVs vary 
for each option but are all of ${\cal{O}}(1\, {\rm to}\, 10)$. 
All other $D$-flat directions generated 
break $F$-flatness at third to fifth order 
in the superpotential.

For the (\ref{sfset}) class of flat directions none of the seven pairs of 
singlets with $Q_Y=\pm \frac{1}{2}$ receive mass. In constrast, 
six pairs must receive FI scale mass for optical unification.  
The three pairs of 
exotic doublets remain FI-scale massless as desired, as do 
the four pairs of exotic triplet/anti-triplets, 
unless $\Phi_2$ receives a VEV. 
In the latter event, one exotic triplet/anti-triplet pair 
receives FI-scale mass from a third order term,
$D_4 \bar{D}_4 <\Phi_2>$, 
and another from a fifth order term,
$D_1 \bar{D}_2 <\Phi_2 \bar{S}_8 S_9 >$.
The VEV of $S_9$ also results in one pair of higgs receiving FI-scale mass
from a third order superpotential term. 
If ${\bar \Phi}_{12}$ or ${\bar \Phi}_{31}$ receive a VEV, a second pair
of higgs also receives similar mass.

The singlet flat-direction class (\ref{sfset}) 
is insufficient because of its lack of 
related mass terms for the six extra pairs of singlets.
Since all other singlet $D$-flat directions examined by us lose 
$F$-flatness before sixth order, 
non-Abelian flat directions appear necessary for realization of 
optical unification in our model.
Thus we have begun to investigate
phenomenologies of flat directions containing VEVs of the model's 
hidden sector fields. The model contains 
four pairs of $SU(5)_H$ $5/\bar{5}$ fields,   
and four pairs of $SU(3)_H$ $3/\bar{3}$ fields. 
We have found that all basis elements of non-Abelian $D$-flat
directions make a positive contribution to the anomalous $D$-term.
Thus, any $D$-flat direction with non-Abelian fields must
contain the singlet direction ${\cal D}_1$ (the only basis element 
with a net negative contribution) with a sufficiently large positive 
coefficient. 
The results of our non-Abelian flat direction study will appear in 
Ref.\ \cite{naoptun}.  We will also be undertaking a
systematic search for additional heterotic string models 
with optical unification potential.

Optical unification offers a phenomenologically appealing resolution
to the factor of twenty difference between 
the apparent $\Lambda_U \sim 2.5 \times 10^{16}$ GeV scale for MSSM 
unification scale, assuming an intermediate scale desert, 
and the lower bound of 
$\Lambda_H \sim 5 \times 10^{17}$ GeV for
coupling unification in weakly coupled heterotic strings. 
However, it is still unclear whether optical unification can, in fact, be
realized in string models.  Whether this is possible for NAHE-based 
free fermionic models, in particular, should be determined 
within the next year.

\section{Acknowledgments}
This paper is a product of the 2003 NSF REU 
program sponsored by the Center for Astrophysics,
Space Physics, and Engineering Research (CASPER) at Baylor University.
Research funding for 
Eric Kasper, Matthew Robinson, and Kristin Stone
was provided by NSF grant no.\  0097386.
G.C. thanks Alon Faraggi and Dimitri Nanopoulos 
for numerous discussions of NAHE-based models and 
thanks Joel Giedt for discussions regarding optical unification.

\newpage
\def\AEF{A.E. Faraggi}
\def\AP#1#2#3{{\it Ann.\ Phys.}\/ {\bf#1} (#2) #3}
\def\JHEP#1#2#3{{\it JHEP}\/ {\bf #1} (#2) #3}
\def\NPB#1#2#3{{\it Nucl.\ Phys.}\/ {\bf B#1} (#2) #3}
\def\NPBPS#1#2#3{{\it Nucl.\ Phys.}\/ {{\bf B} (Proc. Suppl.) {\bf #1}} (#2) 
 #3}
\def\PLB#1#2#3{{\it Phys.\ Lett.}\/ {\bf B#1} (#2) #3}
\def\PRD#1#2#3{{\it Phys.\ Rev.}\/ {\bf D#1} (#2) #3}
\def\PRL#1#2#3{{\it Phys.\ Rev.\ Lett.}\/ {\bf #1} (#2) #3}
\def\PRT#1#2#3{{\it Phys.\ Rep.}\/ {\bf#1} (#2) #3}
\def\PTP#1#2#3{{\it Prog.\ Theo.\ Phys.}\/ {\bf#1} (#2) #3}
\def\MODA#1#2#3{{\it Mod.\ Phys.\ Lett.}\/ {\bf A#1} (#2) #3}
\def\MPLA#1#2#3{{\it Mod.\ Phys.\ Lett.}\/ {\bf A#1} (#2) #3}
\def\IJMP#1#2#3{{\it Int.\ J.\ Mod.\ Phys.}\/ {\bf A#1} (#2) #3}
\def\IJMPA#1#2#3{{\it Int.\ J.\ Mod.\ Phys.}\/ {\bf A#1} (#2) #3}
\def\nuvc#1#2#3{{\it Nuovo Cimento}\/ {\bf #1A} (#2) #3}
\def\RPP#1#2#3{{\it Rept.\ Prog.\ Phys.}\/ {\bf #1} (#2) #3}
\def\etal{{\it et al\/}}
               

\def\bibiteml#1#2{ }
\bibliographystyle{unsrt}

\appendix
\def\s{\phantom{-}}

\begin{table}[t]
\caption{Optical Unification Higgs \& MSSM Exotics\label{tab:one}}
\vspace{0.2cm}
\begin{center}
\footnotesize
\begin{tabular}{|l||c|cccccccc|c|cc|}
\hline
  $F$      & $(SU(3)_C,$ & $Q_{Y}$ & $Q_{Z'}$
   & $Q_A$ & $Q_{1}$ & $Q_{2}$ & $Q_{3}$         & $Q_{4}$
           & $Q_{5}$ & $(SU(5)_{H},$   & $Q_{6}$ & $Q_{7}$  \\
           & $SU(2)_L)$ & & & & & & & & & $SU(3)_{H})$ & &  \\
\hline
      $h_1$& $(1,2)$        &-1/2 & 1/2 & 1  &-1  & 1  & 0 & 0 & 0 & $(1,1)$ & 0 & 0\\
      $h_2$& $(1,2)$        &-1/2 & 1/2 & 1  & 1  & 1  & 0 & 0 & 0 & $(1,1)$ & 0 & 0\\
      $h_3$& $(1,2)$        &-1/2 & 1/2 & 1  & 0  &-2  & 0 & 0 & 0 & $(1,1)$ & 0 & 0\\
      $h_4$& $(1,2)$        &-1/2 & 0   &-1/4&-1/2&1/2 & 0 & 0 & 0 & $(1,1)$ & 2 & 0\\
$\bar{h}_1$& $(1,2)$        & 1/2 &-1/2 &-1  & 1  &-1  & 0 & 0 & 0 & $(1,1)$ & 0 & 0\\
$\bar{h}_2$& $(1,2)$        & 1/2 &-1/2 &-1  &-1  &-1  & 0 & 0 & 0 & $(1,1)$ & 0 & 0\\
$\bar{h}_3$& $(1,2)$        & 1/2 &-1/2 &-1  & 0  & 2  & 0 & 0 & 0 & $(1,1)$ & 0 & 0\\
$\bar{h}_4$& $(1,2)$        & 1/2 & 0   &1/4 &1/2 &-1/2& 0 & 0 & 0 & $(1,1)$ &-2 & 0\\
\hline
      $D_1$& $(3,1)$        &-1/3 &-1/3 & 1  & 0  & 1  & 0  &  0 & 0  & $(1,1)$ & 0 & 0 \\
      $D_2$& $(3,1)$        &-1/3 &-1/3 &-1  & 0  &-1  & 0  &  0 & 0  & $(1,1)$ & 0 & 0 \\
      $D_3$& $(3,1)$        &-1/3 & 1/6 & 1/4&-1/2&-1/2& 0  &  0 & 0  & $(1,1)$ &-2 & 0 \\
      $D_4$& $(3,1)$        & 1/6 & 1/6 & 0  & 0  & 0  &1/2 & 1/2& 1/2& $(1,1)$ &1/2&-15/2\\
$\bar{D}_1$& $({\bar 3},1)$ & 1/3 & 1/3 &-1  & 0  &-1  & 0  &  0 & 0  & $(1,1)$ & 0 & 0 \\
$\bar{D}_2$& $({\bar 3},1)$ & 1/3 & 1/3 & 1  & 0  & 1  & 0  &  0 & 0  & $(1,1)$ & 0 & 0 \\
$\bar{D}_3$& $({\bar 3},1)$ & 1/3 & 1/6 &-1/4&1/2 &1/2 & 0  &  0 & 0  & $(1,1)$ & 2 & 0 \\
$\bar{D}_4$& $({\bar 3},1)$ &-1/6 &-1/6 & 0  & 0  & 0  &-1/2&-1/2&-1/2& $(1,1)$ &-1/2& 15/2\\
\hline
      $X_1$& $(1,2)$& 0   & 0   & 1/2&-1/2& 1/2&1/2& 0  &1/2& $(1,1)$ &-1/2& 15/2\\
      $X_2$& $(1,2)$& 0   & 0   & 1/2& 1/2& 1/2& 0 &-1/2&1/2& $(1,1)$ &-1/2& 15/2\\
      $X_3$& $(1,2)$& 0   & 0   & 1/2& 0  &-1  &1/2&-1/2& 0 & $(1,1)$ &-1/2& 15/2\\
$\bar{X}_1$& $(1,2)$& 0   & 0   &-1/2& 1/2&-1/2&1/2& 0  &1/2& $(1,1)$ & 1/2&-15/2\\
$\bar{X}_2$& $(1,2)$& 0   & 0   &-1/2&-1/2&-1/2& 0 &-1/2&1/2& $(1,1)$ & 1/2&-15/2\\
$\bar{X}_3$& $(1,2)$& 0   & 0   &-1/2& 0  & 1  &1/2&-1/2& 0 & $(1,1)$ & 1/2&-15/2\\
\hline
      $A_1$& $(1,1)$        & 1/2 & 1/2 & 0  & 0  & 0  & 1/2& 1/2&-1/2& $(1,1)$&-1/2& 15/2\\
      $A_2$& $(1,1)$        &-1/2 & 1/2 &-1/2&-1/2&-1/2& 0  & 1/2&-1/2& $(1,1)$&-1/2& 15/2\\
      $A_3$& $(1,1)$        &-1/2 & 1/2 &-1/2& 0  & 1  &-1/2& 1/2& 0  & $(1,1)$&-1/2& 15/2\\
      $A_4$& $(1,1)$        &-1/2 & 1/2 &-1/2& 1/2&-1/2&-1/2& 0  &-1/2& $(1,1)$&-1/2& 15/2\\
      $A_5$& $(1,1)$        & 1/2 &-1/2 &-1/2&-1/2&-1/2& 0  & 1/2&-1/2& $(1,1)$&-1/2& 15/2\\
      $A_6$& $(1,1)$        & 1/2 &-1/2 &-1/2& 0  & 1  &-1/2& 1/2& 0  & $(1,1)$&-1/2& 15/2\\
      $A_7$& $(1,1)$        & 1/2 &-1/2 &-1/2& 1/2&-1/2&-1/2& 0  &-1/2& $(1,1)$&-1/2& 15/2\\
$\bar{A}_1$& $(1,1)$        &-1/2 &-1/2 & 0  & 0  & 0  &-1/2&-1/2& 1/2& $(1,1)$& 1/2&-15/2\\
$\bar{A}_2$& $(1,1)$        & 1/2 &-1/2 & 1/2& 1/2& 1/2& 0  & 1/2&-1/2& $(1,1)$& 1/2&-15/2\\
$\bar{A}_3$& $(1,1)$        & 1/2 &-1/2 & 1/2& 0  &-1  &-1/2& 1/2& 0  & $(1,1)$& 1/2&-15/2\\
$\bar{A}_4$& $(1,1)$        & 1/2 &-1/2 & 1/2&-1/2& 1/2&-1/2& 0  &-1/2& $(1,1)$& 1/2&-15/2\\
$\bar{A}_5$& $(1,1)$        &-1/2 & 1/2 & 1/2& 1/2& 1/2& 0  & 1/2&-1/2& $(1,1)$& 1/2&-15/2\\
$\bar{A}_6$& $(1,1)$        &-1/2 & 1/2 & 1/2& 0  &-1  &-1/2& 1/2& 0  & $(1,1)$& 1/2&-15/2\\
$\bar{A}_7$& $(1,1)$        &-1/2 & 1/2 & 1/2&-1/2& 1/2&-1/2& 0  &-1/2& $(1,1)$& 1/2&-15/2\\
\hline
\end{tabular}
\end{center}
\end{table}

\begin{table}[t]
\caption{Optical Unification Singlets with $Q_Y=0$\label{tab:two}}
\vspace{0.2cm}
\begin{center}
\footnotesize
\begin{tabular}{|l||c|cccccccc|c|cc|}
\hline
  $F$      & $(SU(3)_C,$ & $Q_{Y}$ & $Q_{Z'}$
   & $Q_A$ & $Q_{1}$ & $Q_{2}$ & $Q_{3}$         & $Q_{4}$
           & $Q_{5}$ & $(SU(5)_{H},$   & $Q_{6}$ & $Q_{7}$  \\
           & $SU(2)_L)$ & & & & & & & & & $SU(3)_{H})$ & &  \\
\hline
  $\Phi_1$ & $(1,1)$        & 0   & 0   & 0 & 0 & 0 & 0 & 0 & 0 & $(1,1)$ & 0 & 0\\
  $\Phi_2$ & $(1,1)$        & 0   & 0   & 0 & 0 & 0 & 0 & 0 & 0 & $(1,1)$ & 0 & 0\\
  $\Phi_3$ & $(1,1)$        & 0   & 0   & 0 & 0 & 0 & 0 & 0 & 0 & $(1,1)$ & 0 & 0\\
$\Phi_{12}$& $(1,1)$        & 0   & 0   & 0 &-2 & 0 & 0 & 0 & 0 & $(1,1)$ & 0 & 0\\
$\Phi_{23}$& $(1,1)$        & 0   & 0   & 0 & 1 &-3 & 0 & 0 & 0 & $(1,1)$ & 0 & 0\\
$\Phi_{31}$& $(1,1)$        & 0   & 0   & 0 &-1 &-3 & 0 & 0 & 0 & $(1,1)$ & 0 & 0\\
${\bar\Phi}_{12}$&$(1,1)$   & 0   & 0   & 0 & 2 & 0 & 0 & 0 & 0 & $(1,1)$ & 0 & 0\\
${\bar\Phi}_{23}$&$(1,1)$   & 0   & 0   & 0 &-1 & 3 & 0 & 0 & 0 & $(1,1)$ & 0 & 0\\
${\bar\Phi}_{31}$&$(1,1)$   & 0   & 0   & 0 & 1 & 3 & 0 & 0 & 0 & $(1,1)$ & 0 & 0\\
      $S_1$& $(1,1)$        & 0   & 0   & 0 &-1 & 0 &-1 & 0 & 0 & $(1,1)$ & 0 & 0\\
      $S_2$& $(1,1)$        & 0   & 0   & 0 &-1 & 0 & 1 & 0 & 0 & $(1,1)$ & 0 & 0\\
      $S_3$& $(1,1)$        & 0   & 0   & 0 &-1 & 0 & 0 &-1 & 0 & $(1,1)$ & 0 & 0\\
      $S_4$& $(1,1)$        & 0   & 0   & 0 &-1 & 0 & 0 & 1 & 0 & $(1,1)$ & 0 & 0\\
      $S_5$& $(1,1)$        & 0   & 0   & 0 &-1 & 0 & 0 & 0 &-1 & $(1,1)$ & 0 & 0\\
      $S_6$& $(1,1)$        & 0   & 0   & 0 &-1 & 0 & 0 & 0 & 1 & $(1,1)$ & 0 & 0\\
      $S_7$& $(1,1)$        & 0   & 1/2 &3/4&-1/2&-3/2&0& 0 & 0 & $(1,1)$ & 2 & 0\\
      $S_8$& $(1,1)$        & 0   & 1/2 &3/4&1/2&3/2& 0 & 0 & 0 & $(1,1)$ & 2 & 0\\
      $S_9$& $(1,1)$        & 0   & 1/2 &-5/4&1/2&-1/2&0& 0 & 0 & $(1,1)$ & 2 & 0\\
$\bar{S}_1$& $(1,1)$        & 0   & 0   & 0 & 1 & 0 & 1 & 0 & 0 & $(1,1)$ & 0 & 0\\
$\bar{S}_2$& $(1,1)$        & 0   & 0   & 0 & 1 & 0 &-1 & 0 & 0 & $(1,1)$ & 0 & 0\\
$\bar{S}_3$& $(1,1)$        & 0   & 0   & 0 & 1 & 0 & 0 & 1 & 0 & $(1,1)$ & 0 & 0\\
$\bar{S}_4$& $(1,1)$        & 0   & 0   & 0 & 1 & 0 & 0 &-1 & 0 & $(1,1)$ & 0 & 0\\
$\bar{S}_5$& $(1,1)$        & 0   & 0   & 0 & 1 & 0 & 0 & 0 & 1 & $(1,1)$ & 0 & 0\\
$\bar{S}_6$& $(1,1)$        & 0   & 0   & 0 & 1 & 0 & 0 & 0 &-1 & $(1,1)$ & 0 & 0\\
$\bar{S}_7$& $(1,1)$        & 0   &-1/2 &-3/4&1/2&3/2&0 & 0 & 0 & $(1,1)$ &-2 & 0\\
$\bar{S}_8$& $(1,1)$        & 0   &-1/2 &-3/4 &-1/2&-3/2& 0 & 0 & 0 & $(1,1)$ &-2 & 0\\
$\bar{S}_9$& $(1,1)$        & 0   &-1/2 & 5/4 &-1/2& 1/2& 0 & 0 & 0 & $(1,1)$ &-2 & 0\\
\hline
\end{tabular}
\end{center}
\end{table}

\def\P{\Phi}
\def\p{\phi}

\begin{table}[t]
\caption{$D$-Flat Basis Elements.
The first entry in a given row denotes the $D$-flat basis element, 
the second entry its anomalous charge,
and the remaining entries, the ratios of the squares of the norms of the field VEVs 
defining the basis element.\label{tab:three}}
\vspace{0.2cm}
\begin{center}
\footnotesize
\begin{tabular}{|l||r|rrrrrrrrrrrrrrr|}
\hline
Dir& $Q^{(A)}$& $\P_{12}$&$\P_{23}$&$\P_{31}$& $S_1$& $S_2$& $S_3$
     & $S_4$ & $S_5$ & $S_6$ & $S_7$ & $S_8$ & $S_9$ & $\P_1$ & $\P_2$ & $\P_3$\\
\hline
${\cal D}_1$& -6&    0&    0&    -2&   0&  0&  0& 0&  -1& -1&  0& -3&  3& 0&  0&  0\\
${\cal D}_2$&  0&    0&    0&    -1&   0&  0&  0& 0&  -1& -1&  1& -1&  0& 0&  0&  0\\
${\cal D}_3$&  0&    0&   -1&    -1&   0&  0&  0& 0&  -1& -1&  0&  0&  0& 0&  0&  0\\
${\cal D}_4$&  0&    0&    0&     0&   0&  0&  1& 1&  -1& -1&  0&  0&  0& 0&  0&  0\\
${\cal D}_5$&  0&    0&    0&     0&   1&  1&  0& 0&  -1& -1&  0&  0&  0& 0&  0&  0\\
${\cal D}_6$&  0&   -1&    0&     0&   1&  1&  0& 0&  -1& -1&  0&  0&  0& 0&  0&  0\\
${\cal D}_7$&  0&    0&    0&     0&   0&  0&  0& 0&   0&  0&  0&  0&  0& 1&  0&  0\\
${\cal D}_8$&  0&    0&    0&     0&   0&  0&  0& 0&   0&  0&  0&  0&  0& 0&  1&  0\\
${\cal D}_9$&  0&    0&    0&     0&   0&  0&  0& 0&   0&  0&  0&  0&  0& 0&  0&  1\\
\hline
\end{tabular}
\end{center}
\end{table}

\end{document}